# Simulation of Carbon Nanotube Welding through Ar Bombardment

Mustafa U. Kucukkal and Steven J. Stuart*

Department of Chemistry, Clemson University, Clemson, SC 29634

E-mail: ss@clemson.edu

---

*To whom correspondence should be addressed

Abstract

Single-walled carbon nanotubes show promise as nanoscale transistors, for nanocomputing applications. This use will require appropriate methods for creating electrical connections between distinct nanotubes, analogous to welding of metallic wires at larger length scales, but methods for performing nanoscale chemical welding are not yet sufficiently understood. This study examined the effect of Ar bombardment on the junction of two crossed single-walled carbon nanotubes, to understand the value and limitations of this method for generating connections between nanotubes. A geometric criterion was used to assess the quality of the junctions formed, with the goal of identifying the most productive conditions for experimental ion bombardment. In particular, the effects of nanotube chirality, Ar impact kinetic energy, impact particle flux and fluence, and annealing temperature were considered. The most productive bombardment conditions, leading to the most crosslinking of the tubes with the smallest loss of graphitic (i.e. conductive) character, were found to be at relatively mild impact energies (100 eV), with low flux and high-temperature (3000 K) annealing. Particularly noteworthy for experimental application, a high junction quality is maintained for a relatively broad range of fluences, from $3 \cdot 10^{19}$ m$^{-2}$ to at least $1 \cdot 10^{20}$ m$^{-2}$.

# Introduction

Ever since their discovery by Iijima,[1] carbon nanotubes (CNTs) have been a target of research, due to their electrical and mechanical properties.[2-7] Single-walled carbon nanotubes (SWCNTs), in particular, offer the promise of both extreme and tunable electrical properties, exhibiting promise as nanoelectronic materials. Semiconducting SWCNTs have high carrier mobility[8,9] and band gaps that vary predictably with tube diameter.[10] Metallic SWCNTs, meanwhile, can carry high current densities.[11] These unique properties have driven research to the point where it is realistic to consider computing devices that use nanoscale transistors formed from CNTs i.e CNT field effect transistors (CNTFETs).[3,12,13] The studies about nanoscale transistors formed by CNT junctions, especially Y-shaped intersection of CNTs[14-17] opens a new dimension to CNT usage/application.

Electronic device applications require that SWCNT junctions be formed by creating new covalent bonding connections between the individual tubes, while preserving the desired electrical conductivity at the intersection. A number of different approaches have been considered experimentally for forming these nanotube junctions, including electron beam irradiation (during transmission electron microscopy),[2,6,18] fast atom bombardment,[5,19-21] and "nanosoldering" via local chemical vapor deposition of metal ions.[4,22-24] Experimental methods can provide only limited information on the covalent bonding and chemical structure of the junction, however. Computational methods, such as molecular dynamics, on the other hand, can provide atomic-scale details of the structure and evolution of bonding at the interface between the two nanotubes.

A variety of computational studies have simulated junction formation via bombardment with species ranging from electrons to atomic or molecular ions, or via purely thermal processes. The defects required to induce restructuring at the interface between the CNTs can be caused by noble gas impacts, as illustrated by Krasheninnikov et al., who modeled bombardment with Ar.[19,20,25] Chemical, rather than purely physical, routes to CNT junction formation have also been examined, as in the $CH^+$ ion bombardment modeled by Ni et al.,[26] in which the bombarding particle can assist in forming the covalent bonds. Irradiation with a beam of electrons, one of the most commonly utilized experimental techniques for forming CNT junctions, has been modeled by Jang et al.[27] where impact of electrons to carbon atoms were mimicked by transferring impact energy of 10 keV to randomly chosen primary knock-out atoms (PKAs). And the source of energy for bond reconstruction need not come from impacts, as several studies have illustrated that CNT junctions can be achieved purely by heating, without any form of radiation.[28-32] Computational studies must necessarily make some simplifying approximations, in comparison to experimental conditions, due to constraints on the time and length scales. For example, several of the previous computational studies have used SWCNTs that are rigidly constrained at the system boundary.[19,20,25,27-29] This prevents the desorption of the nanotubes from the surface (which can occur after an impact for the nm-length SWCNTs used in simulation but not the μm-length experimental systems) but also restricts large-amplitude collective motions that might be physically relevant. Some simulations have been performed for unsupported nanotubes,[28,29] even though the material support surface may have important effects on junction formation, either through heat transport or backscattering of the impacting particles. Most simulations use thermostats to control the system temperature, since impact energies of hundreds of eV can raise the system temperature by hundreds of de- grees Kelvin when thermalized over a system size of only a few thousand atoms. However, these thermostats artificially perturb the forces and/or velocities, which can result in nonphysical bond formation and dissociation when applied during the collision cascade following impact.[28-32] One key limitation of all computational studies is the high flux of impacting particles, which can be orders of magnitude higher than experimental fluxes.[27]

In the present study, we model a SWCNT junction formed by Ar bombardment of two sup- ported orthogonal nanotubes, with the aim of determining the most effective bombardment condi- tions, while improving on some of the limitations of prior simulations. Multiple sequential colli- sions were performed, up to total fluences of $1·10^{20}$ m$^{-2}$. The collisions were performed in the microcanonical ensemble, to avoid any nonphysical effects of the thermostat forces on the reaction dynamics, while the systems were equilibrated at elevated temperature for a separate annealing period after the short-term collision dynamics, in an attempt to decrease the effective flux of the bombardment. The effects of SWCNT structure, impact kinetic energy, annealing temperaure, and annealing time on the quality of the resulting SWCNT junction were evaluated to determine the conditions that lead to the most crosslinking with the least unnecessary damage to the interface.

## Methods

The carbon-carbon interactions were modeled using the adaptive intermolecular reactive empirical bond-order (AIREBO) potential.[33,34] This is a classical bond-order potential of the same general form as the Tersoff[35] and Brenner[36,37] potentials, and can thus model the covalent bond-breaking and bond-forming reactions that are crucial in simulating the bond rearrangements involved in SWCNT junction formation. Unlike these other bond-order models, the AIREBO potential also models non-covalent van der Waals interactions in an adaptive way that includes these interactions between nonbonded atoms, as is required for accurate modeling of the intertube interactions, but allows them to be adaptively switched off (or on) as new covalent bonds are formed (or dissociate). The AIREBO potential has been used successfully to model many carbon-based nanomaterials, including nanotubes.[25,33,38,39]

The van der Waals interactions between the Ar atoms and the C atoms in the SWCNTs were modeled with Lennard-Jones 12-6 interactions, using $\varepsilon_{ArC}/k = 54.507$ K and $\sigma_{ArC} = 3.215$ Å. These interactions were not treated adaptively, since there is no covalent bonding interaction between Ar and C.

Two different SWCNT structures were used in this study, in an attempt to examine the effect of tube structure on junction quality. The (10,10) and (11,9) tubes were chosen for this purpose because they are readily synthesized and widely used in experiments.[40-42] The (10,10) tube is metallic, while the (11,9) tube is semiconducting, so these tubes also allow a study of metallic-metallic, semiconducting-semiconducting, and metallic-semiconducting connections that are of interest for nanoelectronics applications. Additionally, these tubes have nearly the same radius ($r_{(10,10)} = 1.356$ Å and $r_{(11,9)} = 1.358$ Å), which avoids confounding effects from different radii of curvature when the tubes are varied. The (10,10) tube contained 2860 C atoms and had a length of 17.35 nm, whereas the (11,9) tube contained 2855 atoms and was 17.38 nm long. A total of three different junctions were prepared from crossing these two SWCNT types: the homogeneous junction of (10,10) tubes, homogeneous junction of (11,9) tubes, and the heterogeneous junction of a (10,10) tube with an (11,9) tube where the (10,10) tube is closer to the substrate.

Although free-standing tubes can be constructed,Including this substrate support in the model is important, as it has effects on the geometry of the adsorbed tubes, the dynamics of the impacting ion (which may either penetrate or scatter off the substrate) and the post-collision dynamics of the nanotubes. In order to reduce the computational cost of the simulation, we model the substrate with a continuum model, rather than a fully atomistic model, while still allowing for energy-dependent penetration or back-scattering of the impacting ion.

For the Lennard-Jones (LJ) interaction of a single atom with the continuum substrate model, we use the LJ 9-3 potential, which is obtained by integrating the usual LJ 12-6 potential,

$$E_{LJ}(r) = 4\varepsilon \left[ \left(\frac{\sigma}{r}\right)^{12} - \left(\frac{\sigma}{r}\right)^{6} \right], \quad (1)$$

over a continuous distribution of particles with number density $\rho$ in the half-space $z \leq 0$ interacting with a particle at an arbitrary position $(x, y, z)$ with $z > 0$, yielding

$$E_{LJS}(z) = -8\pi\rho\varepsilon\sigma^3 z^{-6} \left[ \frac{\sigma^6}{90 z^6} - \frac{1}{12} \right]. \quad (2)$$

The parameters $\sigma$, $\varepsilon$, and $\rho$ can be chosen to represent any desired substrate material. Here, we choose values representative of a graphite support, with $\rho = 0.11$ Å$^{-3}$ (corresponding to a density of 2.2 g/cm$^3$), $\sigma = 3.4$ Å, and $\varepsilon/k = 24.8$ K; the potential is active for $z \leq 9.0\sigma$, and is switched off smoothly by $z = 12.0\sigma$ using a cubic spline. The choice of graphite as a substrate material is fairly arbitrary. We assume that the substrate parameterization does not have a significant impact on the junction formation, beyond the simple presence of a support surface to induce some structural deformation in the soft nanotubes, interact with the collisionally excited tubes after impact, and scatter ejected particles.

The scattering of energetic particles by the substrate is an important factor in accurate modeling of the chemical reorganization at the interface, however. Both the impacting Ar atom as well as ejected C atoms can be backscattered from the substrate, providing an additional channel for generating chemical changes in the nanotubes. Neglecting this effect can lead to underestimation of the degree of restructuring caused at a given fluence. Modeling the substrate as an idealized LJ surface, however, with an infinite repulsive wall, will elastically reflect 100% of the ejected atoms, regardless of kinetic energy, and this is also not realistic. In a physical system, some fraction of the ejected atoms will penetrate the surface, and the fraction of reflected atoms decreases to near 0% at high kinetic energies. Based on experimental results[43] indicating that the penetration threshold for Ar$^+$ impacts on graphite is 43.5 eV, we chose to truncate the repulsive wall at a height of 50 eV, so that the substrate interaction potential used is max (50 eV, $E_{LJS}$). Impacting particles with a kinetic energy greater than 50 eV will thus penetrate the surface rather than be reflected. In the simulations, any atoms (C or Ar) with $z < 0$ and $v_z < 0$ that have penetrated the surface are removed from the system, rather than continuing to integrate their dynamics. To construct the system, one SWCNT is aligned parallel to the plane of the substrate, with its lowermost atoms placed 2.9 Å above the implicit surface. The second (upper) SWCNT is then placed with its long axis perpendicular to that of the first (lower) SWCNT, and parallel to the plane of the substrate. If this upper tube is initialized too far from the substrate, beyond the finite range of the switched LJ potential, there will be no force inducing it to adsorb to the substrate. Consequently, we generated an initial configuration by deforming the upper SWCNT so that its lowermost atoms were 2.9 Å above the implicit surface for most of its length, but deviated upward in the region where it crosses the lower SWCNT, leaving at least 3.4 Å vertical spacing between the two tubes where they cross. This initial configuration is illustrated in Figure Figure 1. This initial geometry was then equilibrated for 100 ps at a temperature of 3000 K in the canonical ensemble, in order to allow the SWCNTs to relax. The resulting geometry exhibits deformations from the originally cylindrical SWCNT tube geometry, reflecting a compromise between increasing the favorable van der Waals interactions by maximizing tube-tube and tube-substrate contact, while decreasing unfavorable strain energy. This relaxed geometry is illustrated in Figure Figure 1. The reduction of strain energy causes the axis of the upper SWCNT to be nearly linear as it tents over the lower SWCNT, so that its geometry is perturbed by the lower SWCNT for a distance of roughly 80 Å away from the lower SWCNT. The tube lengths of 173.5 Å and 173.8 Å were chosen to ensure that the ends of the upper SWCNT are parallel to the supporting surface, away from the perturbing influence at the junction.

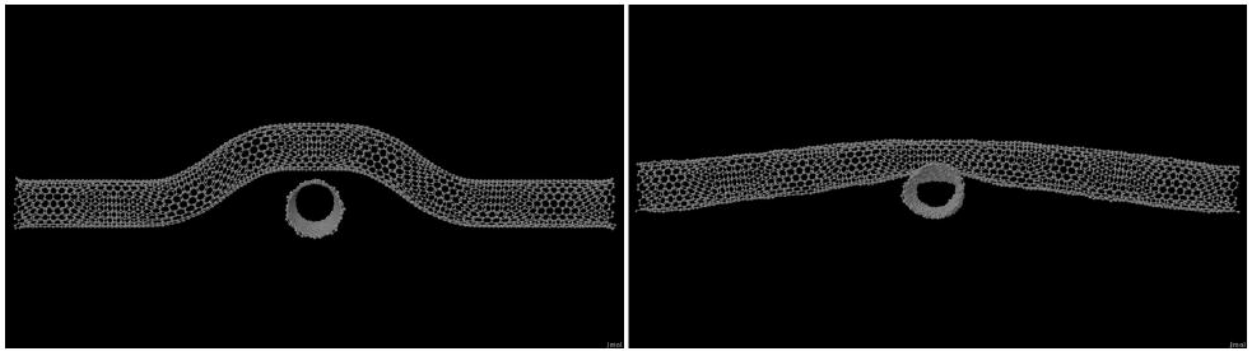

Figure 1: Configuration of a pair of (11,9) SWCNTs before (left panel) and after (right panel) equilibration for 100 ps at 3000 K.

The total adhesive van der Waals interaction between the (11,9) SWCNT of length 173.8 Å and the implicitly modeled substrate is 26.14 eV. Note that this is considerably less than the impact energies of 100 eV or more. Consequently, if a large enough fraction of the impact kinetic energy is transferred to vertical motion of the SWCNT center of mass, a single impact can result in des- orption of the entire SWCNT. This would not happen in experiment, because the physical tubes are much longer, typically several μm in length, with correspondingly larger adsorption energies. Thus, to prevent unphysical impact-induced desorption, one C atom at each end of each of the two SWCNTs (i.e. four atoms in total) are rigidly constrained during equilibration and bombardment. This minimal constraint prevents tube desorption, while still allowing more conformational flex- ibility than the approach of rigidly constraining hundreds of atoms in entire cylindrical collars at

the ends of the tubes, as has been done in some previous simulations.[20,27]

The bombardment is performed by directing an Ar atom towards the junction of the SWCNTs. This junction is located by projecting all carbon atoms into a plane parallel to the substrate surface, and identifying a square region of area 10 Å · 10 Å in which the number density of carbons is the highest. This is done before each impact, allowing for the possibility that the junction location may drift, due to stresses or bonding changes in the SWCNTs. The Ar atom is inserted 10 Å above the uppermost atoms of the top SWCNT, with a randomly chosen lateral position, uniformly distributed in the 100 Å$^2$ impact area. The Ar atom is given a velocity that corresponds to the desired impact kinetic energy, directed perpendicular to the interface. These kinetic energies ranged from 100 eV to 2000 eV in different simulations.

The dynamics are then integrated in the microcanonical ensemble for 1.5 ps using a variable-timestep integrator[44] with a target energy diffusion value of 0.05 eV$^2$/ps. This 1.5 ps period allows enough time for the initial impact and the subsequent collision cascade. The variable-timestep integrator improves the efficiency of the simulations considerably, as the timesteps required during the initial impacts are much smaller than those that can be used to achieve the same level of energy conservation during the later stages after the energy has been dispersed. This is particularly true for large impact energies, where small timesteps are needed to integrate the high-energy primary and secondary collisions. For example, for 2000 eV impacts, the average timestep needed to maintain energy conservation errors below 0.05 $eV^2/ps$ is as small as $1 \cdot 10^{-4}$ fs during the collision, but 0.4 fs after the collision energy has been dissipated into thermal energy.

Performing sequential impacts every 1.5 ps in a region of area 100 $Å^2$ corresponds to a flux of $6.7 \cdot 10^{29}$ $m^{-2} s^{-1}$, many orders of magnitude higher than experimental values. To reduce the effective flux, the system is annealed for between 4 and 8 ps between impacts in the canonical ensemble, at an annealing temperature of between 1000 and 3000 K. This decreases the nominal flux somewhat, to $1.1$-$1.8 \cdot 10^{29}$ $m^{-2} s^{-1}$, but more importantly, the annealing at elevated temperature allows for some degree of structural relaxation, as would occur at much longer times in the period between impacts. Thus the effective flux, while not known exactly, is considerably lower than the nominal flux. The canonical-ensemble annealing is performed using the Langevin thermostat,[45] with a time constant of 100 fs, and a constant timestep of 0.1 fs; variable timesteps are not required because the dynamics are no longer exhibiting fast non-equilibrium relaxation. Any atoms that penetrate the surface are removed from the system (as described earlier), along with reflected atoms (C or Ar) that reach a height of $z = 50$ above the surface with $v_z > 0$.

Up to 100 successive bombardments were performed, each separated by the 1.5 ps microcanonical + 4-8 ps annealing period, for a total fluence of up to $1.0 \cdot 10^{20}$ $m^{-2}$. This is more than a factor of 10 larger than the fluences considered in previous studies,[19,20,27] and many have considered only single impacts. Bombardment conditions were varied by considering impact energies of 100, 500, 1000, 1500, or 2000 eV; annealing times of 4 or 8 ps; and annealing temperatures of 1000, 1500, 2000, or 3000 K. For each set of impact conditions, five independent simulations were carried out, allowing statistical analysis of the data.

We use three different properties to quantify the structural changes at the SWCNT junction, based on the expectation that an ideal junction should have a large number of covalent connec-tions between the two initially distinct tubes, the electrical conductance of this junction should be large, and the tubes should have relatively little damage. Rather than measuring the electrical conductance directly, as has been done in some prior studies,? we use structural proxies for these quantities.

Recognizing that electrical conductivity will require a mostly graphitic structure of the reconstructed interface, we choose our measure of damage to be the loss in the number of $sp^2$ carbons,

$$\Delta n_{sp^2}(\phi) = n_{sp^2}(0) - n_{sp^2}(\phi), \tag{3}$$

where $n_{sp^2}(\phi)$ is the number of $sp^2$-hybridized carbons (i.e. those with a coordination number of 3) in the system at a fluence of $\phi$, measured after the annealing period. The choice of sign ensures that $\Delta n_{sp^2}$ will be positive quantity when $sp^2$ carbons are lost. Note that with this measure a simple crosslink between the two SWCNTs, which converts two $sp^2$ carbons to $sp^3$ will be recognized as damage, because the $sp^3$ carbons will not provide any (substantive) electrical connection. We also recognize that counting $sp^2$ carbons is not sufficient to estimate the electrical conductance; quantum mechanics must be included at some level to estimate conductance from the bonding structure of a carbonaceous system. We treat this quantity as only a heuristic estimate of the damage to the tubes.

As our measure of the degree of connection between the two nanotubes, we use a direct count of the number of crosslinks between the tubes, $n_{cl}$, defined as the number of covalent bonds $C_i$-$C_j$ between an atom i that was originally part of the lower SWCNT and an atom j that was originally part of the upper SWCNT, and in which atom i is additionally bonded to at least 2 other atoms of the upper SWCNT and atom j is bonded to at least 2 other atoms of the lower SWCNT. The latter restrictions are included to ensure that an atom ejected from one nanotube and captured by the other is not erroneously considered as a crosslink. This quantity is also tracked as a function of the fluence, $n_{cl}(\phi)$, and is always measured at the end of the annealing period.

Because a strong, ideal connection will have a large number of crosslimks, but with the minimal amount of damage, we also define a junction quality measure, R, as the ratio of these quantities,

$$R(\phi) = \frac{\overline{\Delta n_{cl}(\phi)}}{n_{sp^2}(\phi)} \quad (4)$$

Note that simple crosslinks are characterized by a value of $R = 0.5$, since two $sp^2$ carbons have been lost (converted to $sp^3$) in the formation of one crosslink. Additional reconstruction of crosslinks to form a more graphitic, $sp^2$-hybridized connection could increase the value of R above 0.5 Thus we can roughly expect that values of $R < 0.5$ indicate junctions with damage in excess of that required to form the connections; values of $R \approx 0.5$ are crosslinked with minimal excess damage, and values of $R > 0.5$ have reconstructed to heal some of the non-graphitic damage.

## Results

First we examine the effect of SWCNT structure on junction quality. Figure 2 shows the difference in quality for the junctions formed by homogeneous pairing of (11,9) tubes and the homogeneous pairing of (10,10) tubes, for 100 eV Ar impacts annealed for 8 ps at 3000 K between impacts and Figure 3 shows the comparison of junction quality measure formed by homegenous pairing of (11,9) and homegenous pairing of (10,10) tubes bombarded with Ar atoms having different kinetic energy starting from 500 eV up to 2000 eV where the tubes are annealed at 3000 K for 8 ps be- tween consecutive impacts. For both tubes, the junction quality, R, increases at low fluences up to $0.4 \cdot 10^{20}$ m$^{-2}$, and then reaches a plateau, remaining roughly constant up to the highest fluences obtained, $1 \cdot 10^{20}$ m$^{-2}$. This same behavior is fairly general for all impact energies and annealing conditions examined, although the rise time and junction quality at the plateau differ under different conditions. Figure 1 through Figure 6 in the Supplemental Information illustrate the evo- lution of junction quality under bombardment for all tube geometries, impact energies, annealing temperatures, and annealing times examined.

Under continued bombardment, the junction quality would surely deteriorate, as the impacts erode would eventually erode away all of the connections in the bombardment region at high fluences. No sustained decrease in junction quality was observed at fluences of $1\cdot 10^{20}$ m$^{-2}$ in any of the systems examined, however. This is encouraging news for experimental junction formation, as it suggests that the junction quality is not very sensitive to fluence, and there is a relatively broad range of fluences at which reasonable junctions may be formed.

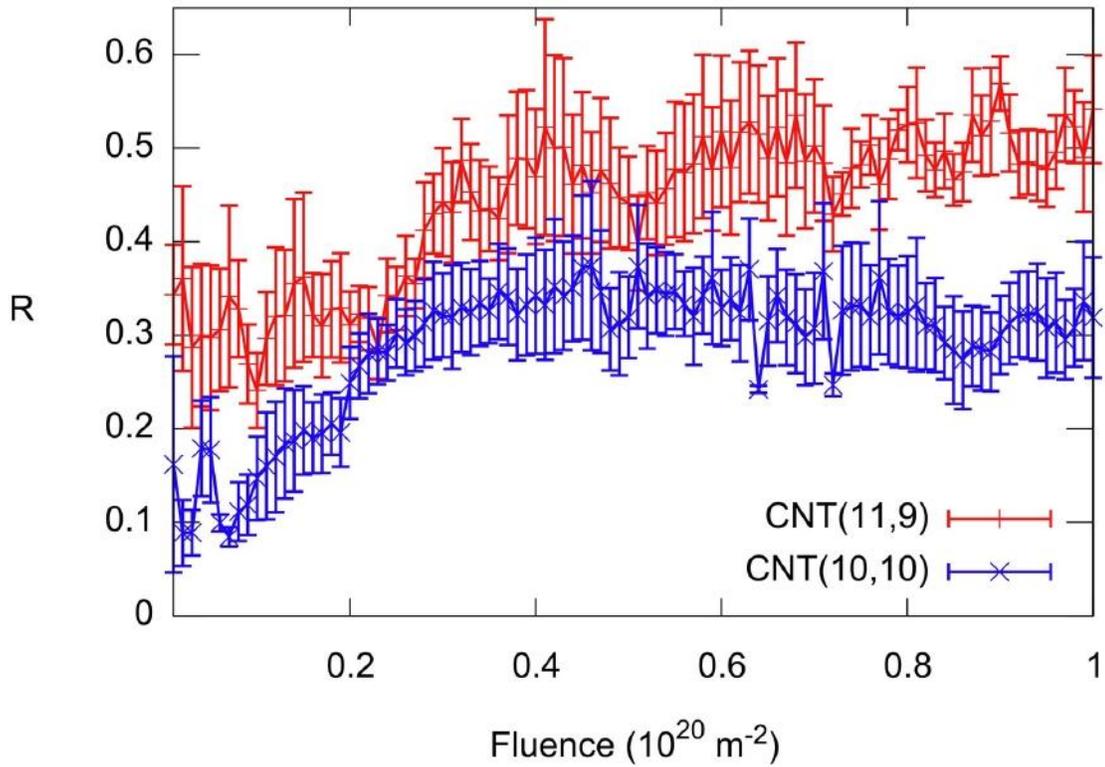

Figure 2: Evolution of junction quality with fluence for junctions formed by two (11,9) SWCNTs (red curve) and two (10,10) SWCNTs (blue curve), under bombardment by 100 eV Ar with 8 ps annealing at 3000 K. Error bars represent standard deviations.

Figure 2 and Figure 3 clearly illustrate that the junction quality is higher for the pair of crossed (11,9) SWCNTs than for the pair of crossed (10,10) SWCNTs. The pair of (10,10) tubes reach a plateau value near $R = 0.3$, indicating that the tubes are damaged more than needed to form the observed number of crosslinks. The pair of (11,9) tubes reach a plateau value near or slightly above $R = 0.5$, however, either because less unnecessary damage was produced, or because this damage was annealed away more successfully. The superior junction quality for the homogeneous

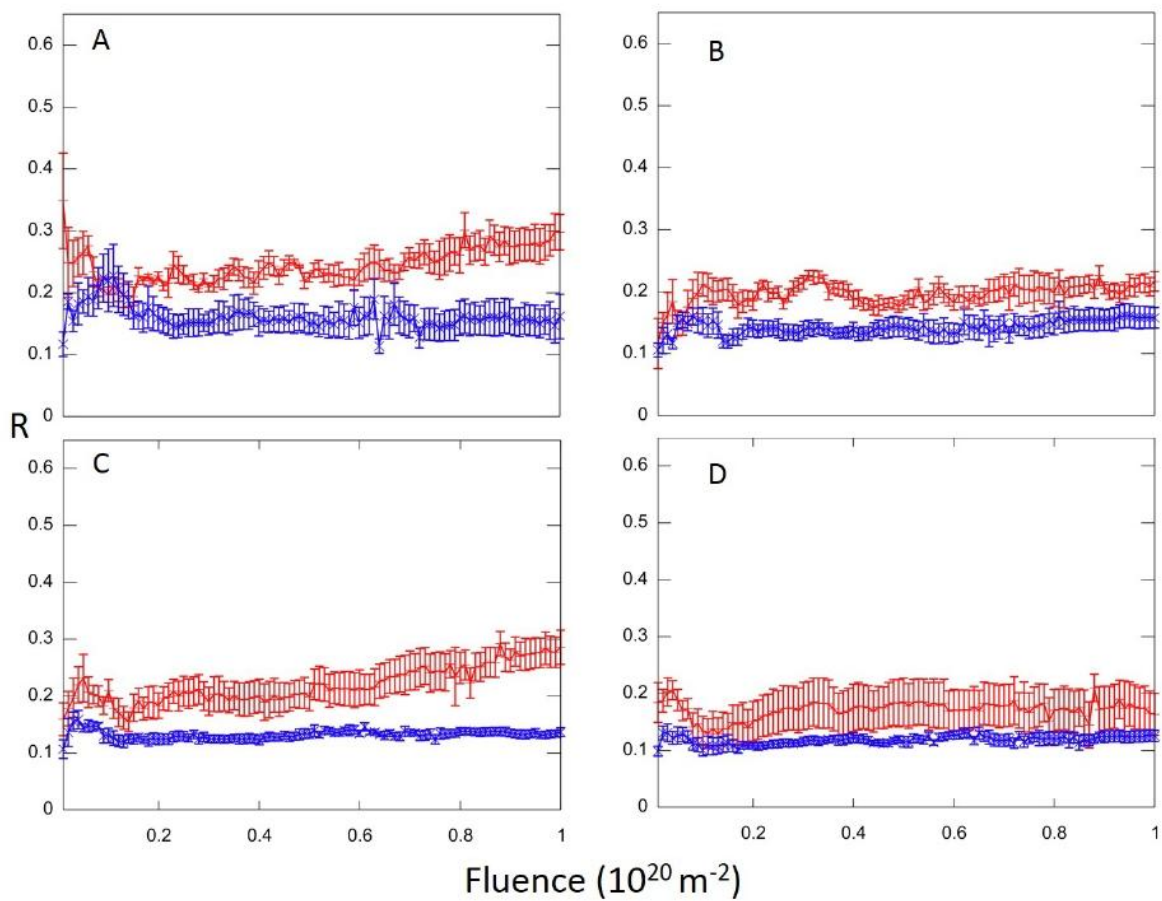

Figure 3: Evolution of junction quality with fluence for junctions formed by two (11,9) SWC- NTs (red curve) and two (10,10) SWCNTs (blue curve), under bombardment by (A) 500 eV, (B)1000 eV, (C)1500 eV and (D)2000 eV Ar atoms with 8 ps annealing at 3000 K. Error bars rep- resent standard deviations.

(11,9) junction is more than anecdotal, and persists across multiple independent simulations, and at a wide range of fluences, as indicated by the error bars in Figure 2. The reason, however, is not clear. Nanotubes of different curvature have different reactivity, but the (11,9) and (10,10) SWCNTs differ in diameter by only 0.1%,[46] so this is unlikely to be the origin of the difference. The (11,9) tube is semiconducting and the (10,10) tube is metallic; this difference in conductivity could result in different reactivities under experimental conditions, particularly under applied fields or in the presence of excess charge. But the simulations include neither quantum mechanical treatment of the electrons, nor electrostatic effects of excess charge, so these factors do not explain the difference in junction formation in the present simulations. The chirality (i.e. handedness and asymmetry) of the (11,9) tube is another difference, compared to the achiral (10,10) tube. The most plausible explanation in the significant diffence of junction quality formed by (11,9) and (10,10) homogenous pair of tubes is the difference in chirality of the tubes. Turning now to the influence of different Ar initial energies, Figure 4 shows how the junction quality measure changes with the fluence for different initial Ar kinetic energies (different colors). The systems were annealed for 8 ps at 3000 K between each bombardment. The junction quality measure is highest ( 0.5) with the lowest initial Ar energy, i.e. 100 eV. Lower ratios obtained with higher initial energies, which incidates that higher energies cause more destruction, i.e. more loss in number of

$sp^2$ hybridized carbon atoms as opposed to forming new connections. Also, based on the visual inspection of results presented in Figure 4, the CNT(11,9) system for all the impact energies reach a steady state fluence value between $0.1 \cdot 10^{20}$ and $0.4 \cdot 10^{20}$ $m^{-2}$ meaning that the ratio does not change significantly with more bombardments. The visual inspection of the trajectories shows that an initial threshold number of bombardments causes damage with a large hole in the impact zone, where the subsequent bombardments mainly passes through without causing any additional change in the junction quality measure. As an example, the extended damage by Ar atoms having impact energy of 2000 eV to CNT(11,9) system annealed at 3000K for 8ps is shown in Figure 5 which produced a junction quality measure of 0.2. The structural change of CNT(11,9) system bombarded with Ar atoms having 100 eV impact energy is shown in Figure 6 with a junction quality measure of 0.5. That is why the zone is not affected significantly after the threshold fluence is reached. As one may expect, the system with the lowest initial Ar energy (100 eV) needed fluence value higher than $0.25 \cdot 10^{20}$ $m^{-2}$ to reach the steady state as opposed to other systems where much less number of bombardments needed to reach the steady state.

In addition, these simulations with varying initial Ar energies were repeated with an annealing time of 4 ps between each bombardment and also with different annealing temperatures of 1000, 1500 and 2000 K. The 100 eV initial Ar energy consistently produced the highest ratio as seen in Figure 8. After determining the optimum energy value (100 eV) that produces the highest junc- tion quality measure, we further investigated the effects of annealing temperature and annealing

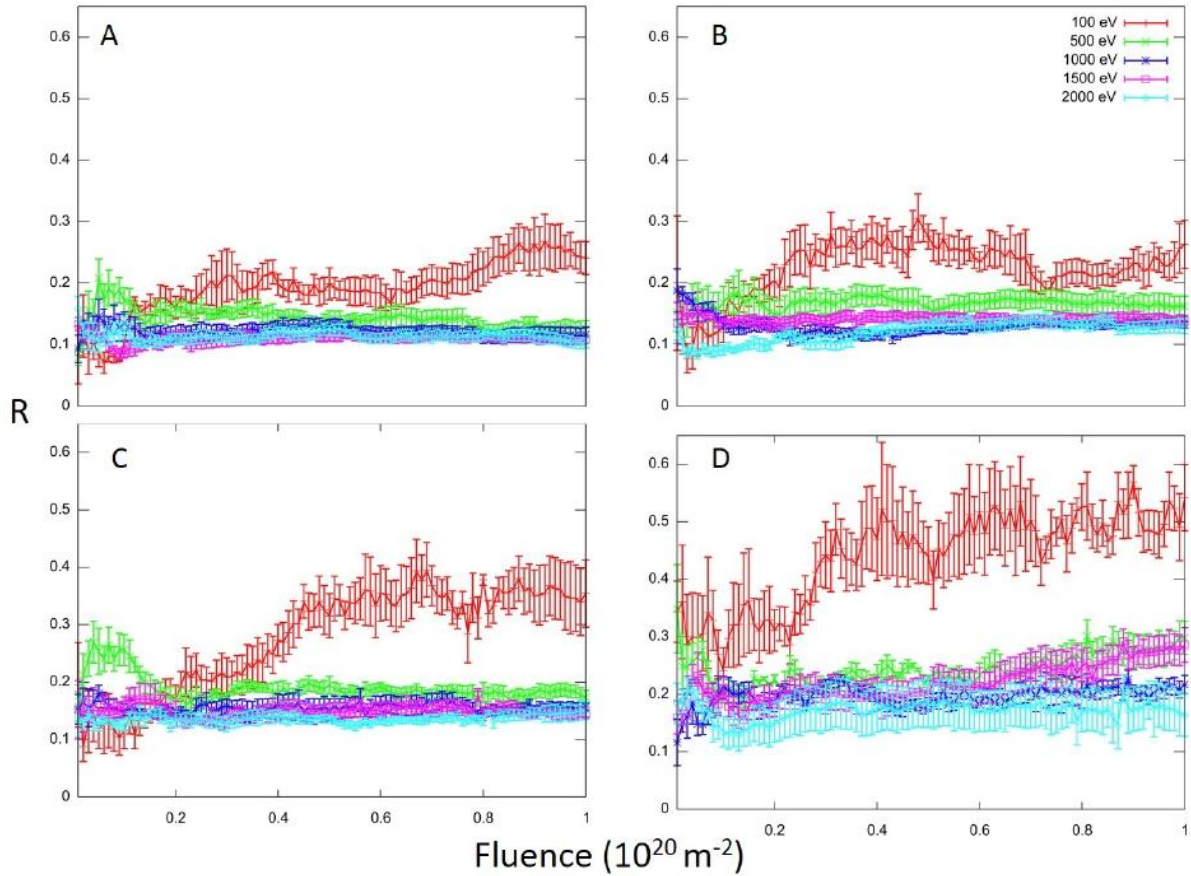

Figure 4: Evolution of junction quality with fluence for junctions formed by (11,9) SWCNTs under bombardment by Ar having various impact energies with 8 ps annealing at (A) 1000 K, (B) 1000 K , (C) 2000 K and (D) 3000 K. Error bars represent standard deviations

time in between bombardments to junction quality measure. The results for CNT(11,9) system at annealing temperatures of 1000 K, 1500 K , 2000 K and 3000 K is given in Figure 4. The CNT(11,9) system annealed at different temperatures reached steady state fluence values between $0.24 \cdot 10^{20}$ and $0.4 \cdot 10^{20}$ m$^{-2}$. These studies also did not reach high fluence value where the junction quality measure is expected to decrease due to the destruction caused by Ar bombardment. The highest junction quality measure achieved increased with increasing annealing temperature. The system annealed at 3000 K produced the highest junction quality measure among the same systems annealed at different temperatures. The junction quality measure for the CNT(11,9) sys- tem annealed at 3000 K at the steady state region is around expected value of 0.5. Increasing the temperature above 3000 K is not practical since temperatures above 3000 K may cause the melting of CNTs. Next, the 3000 K annealing system was selected to investigate the influence of annealing time. Figure 8 shows the results for the CNT(11,9) system with annealing times of 4 ps after Ar bombardment with various impact energies at 3000 K. The same trend was observed as the impact energy increases the damage increases substainly resulting junction quality measure to decrease. The comparison of junction quality measure for (11,9) tubes bombarded with 100 eV Ar atom followed by annealing at 3000 K for 4 ps and 8ps is given in Figure 9. The effect of annealing time was found to be less pronounced although measurable compared to other factors such as the initial Ar energy and annealing temperature. Annealing for 8 ps after each Ar bombardment gives a higher junction quality measure than the one annealed for 4ps, which suggest that increasing the annealing time will lead to more organizations in the junction area after the bombardment.

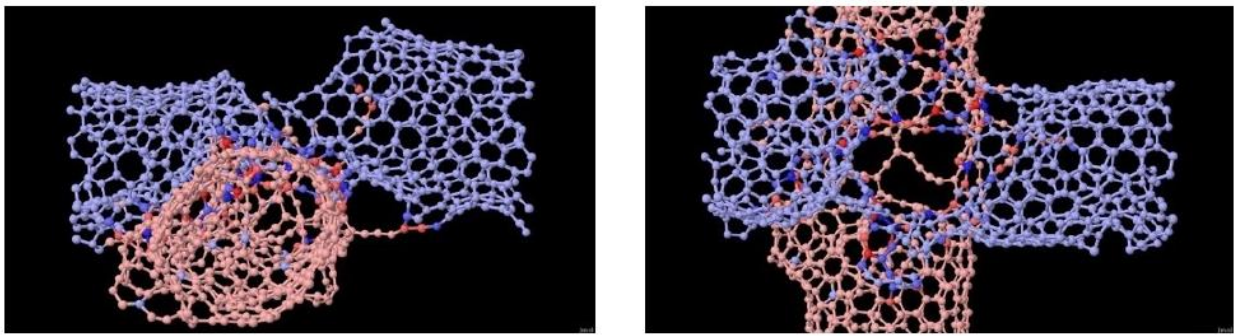

Figure 5: Snapshots of CNT(11,9) system bombarded with Ar atoms having 2000 eV impact energy after annealed at 3000 K for 8 ps (Left) : Front (Right) : Top view (For clarity, long arms are omitted) which produced junction quality measure of 0.2.

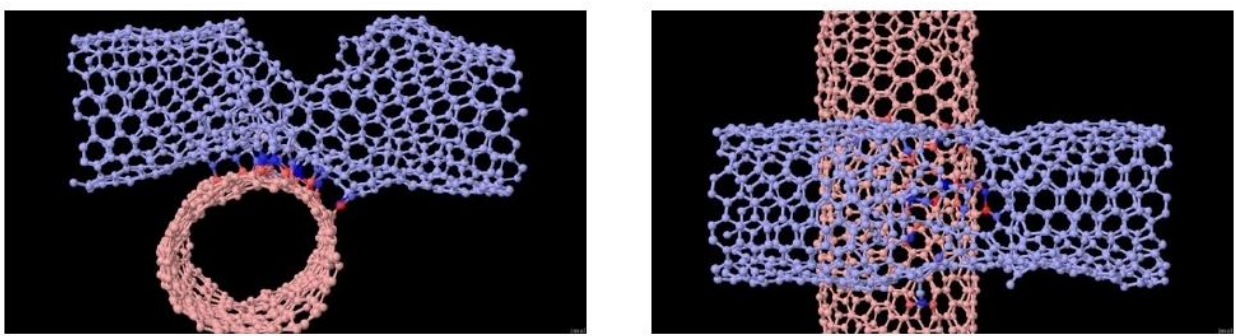

Figure 6: Snapshots of CNT(11,9) system bombarded with Ar atoms having 100 eV impact energy after annealed at 3000 K for 8 ps (Left) : Front (Right) : Top view (For clarity, long arms are omitted) which produced junction quality measure of 0.5.

# Conclusion

We have simulated Ar bombardment on the junction of two SWCNTs placed perpendicular on to one another supported by implicit LJ surface. We investigated the effects of kinetic energy of bombarding atoms, type of the CNTs, annealing temperature and time to the number of connections and the destruction in the structure of SWCNTs. We have found that the number of connections between two SWCNTs increase with the increasing the kinetic energy, however the destruction in the SWCNTs structure inreases with the kinetic energy more than that of the connections number. This causes 100 eV to to be best choice for the kinetic energy of bombarding Ar atoms which leads a reasonable ratio of connections to the lost $sp^2$ C atoms. We concluded that the destruction of SWCNTs decreases with the increasing temperature especially at 100 eV kinetic energy, the damage is significantly the lowest at annealing temperature of 3000 K. It is also found that increasing temperature leads increase in the connections of between SWCNTs. It is assumed higher temperature helps bonding rearrangements to pass the activation barrier. For all the kinetic energies studies, 3000 K becomes the best annealing temperature that gives the best ratio of the connection numbers to the number of loss $sp^2$ C atoms. We also found that optimum annealing time to be 4 ps. Annealing time larger than 4 ps leads more bonding and breaking arrangements in the systems therefore the number of lost $sp^2$ C atoms is more than that of 4ps annealing times also the number of connections is less than that of 4ps annealing time. The networks between SWCNTs can be formed through atom bombardment using current experimental techniques by forming connections in the junction areas. Hopefully, the entangled SCNTs can be used as nano-electronic devices such as transistor with multi-terminal junctions.

# Acknowledgement

The authors gratefully acknowledge computational support by the Clemson CITI staff.

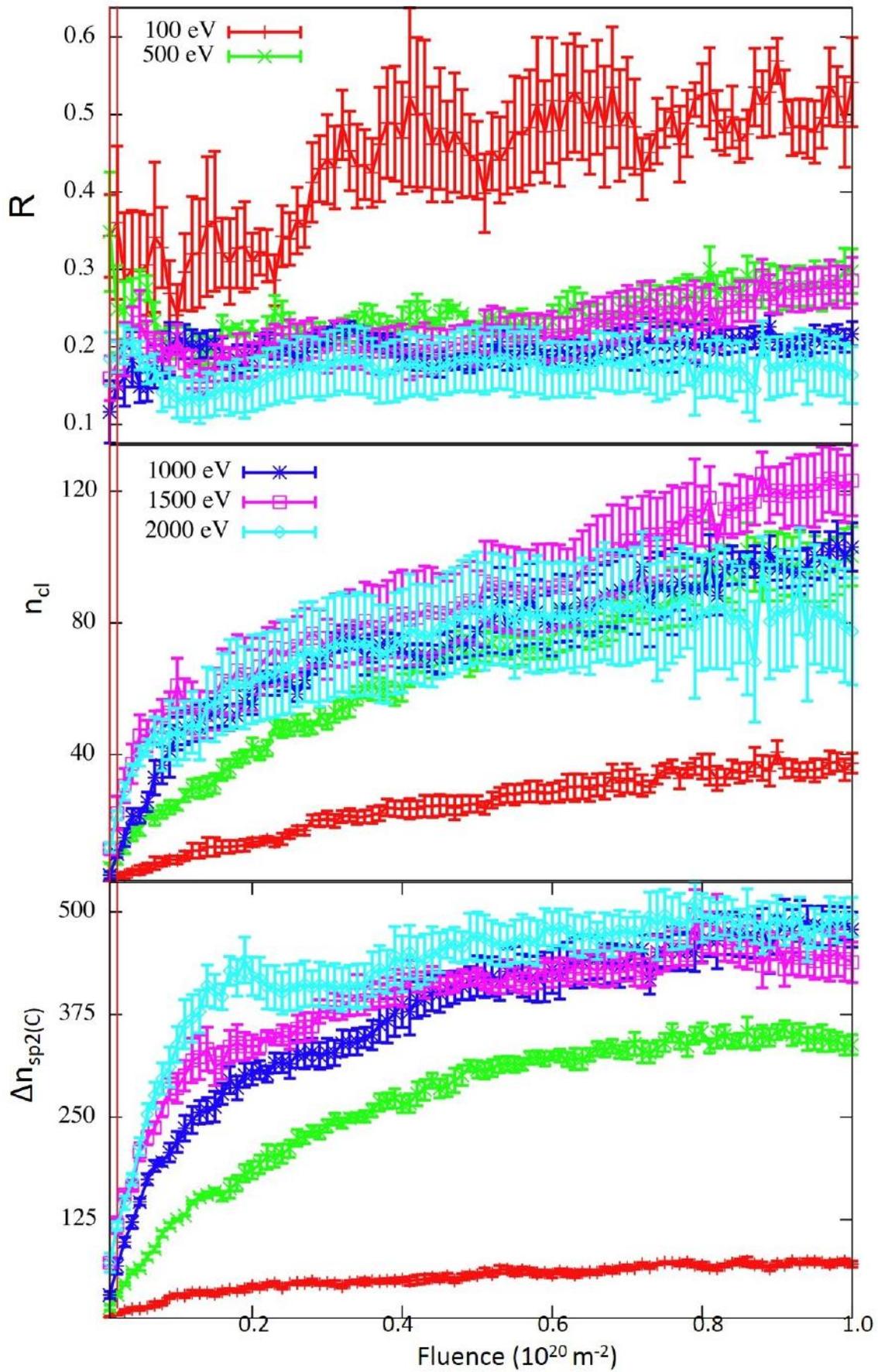

Figure 7: Evolution of junction quality (R), the number of junctions ($n_{cl}$) and the number of sp² C

atoms lost ($\Delta n_l (C)$) with fluence for (11,9) SWCNTs under bombardment by Ar having various

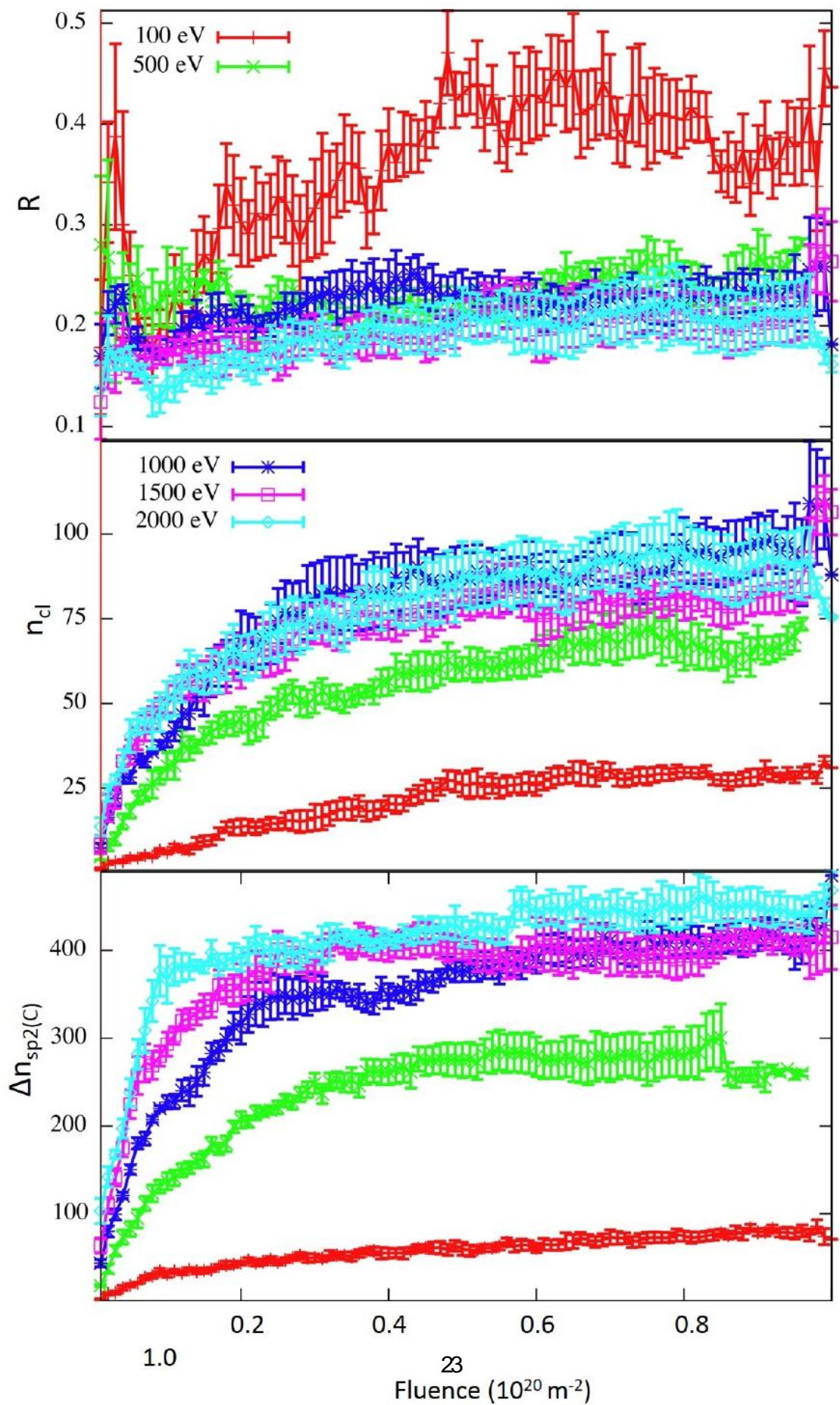

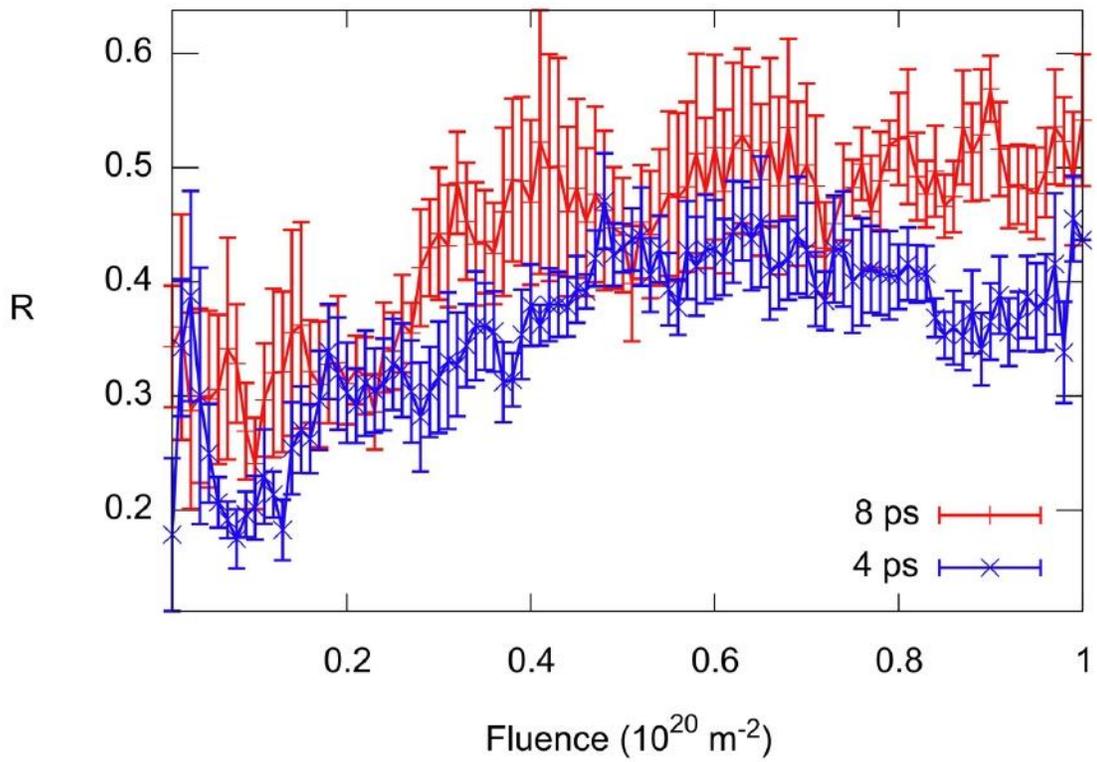

Figure 9: Effect of fluence on R for the CNT(11,9) system annealed at 3000 K for 4ps and 8ps after impacted with Ar atom of 100 eV kinetic energy

24